\documentclass[conference]{IEEEtran}
    \IEEEoverridecommandlockouts
    \pdfoutput=1
    \usepackage{cite}
    \usepackage{amsmath,amssymb,amsfonts}
    \usepackage{algorithmic}
    \usepackage{graphicx}
    \usepackage{textcomp}
    \usepackage{xcolor}
    \usepackage{graphicx}
    \usepackage{multirow}
    \usepackage{url}
    \usepackage{times}
    \usepackage{float}
    \usepackage{balance}
    \usepackage{verbatim}
    \usepackage{subfigure}
    \graphicspath{{./}{figure/}}

    \def\BibTeX{{\rm B\kern-.05em{\sc i\kern-.025em b}\kern-.08em
        T\kern-.1667em\lower.7ex\hbox{E}\kern-.125emX}}
    
\begin{document}

    \title{
    Predicting Channel Quality Indicators for 5G Downlink Scheduling in a Deep Learning Approach
    }

\author{\IEEEauthorblockN{Hao Yin$^*$, Xiaojun Guo$^\dagger$,Pengyu Liu$^\dagger$,  Xiaojun Hei$^\dagger$,Yayu Gao$^\dagger$}
\IEEEauthorblockA{{$^*$Department of Electrical and Computer Engineering, University of Washington} \\ {$^\dagger$School of Electronic Information and Communication, Huazhong University of Science and Technology} \\
Email: $^*$\{haoyin\}@uw.edu,  $^\dagger$\{guoxj, eic\_lpy, heixj, yayugao\}@hust.edu.cn } 
}

%    \author{\IEEEauthorblockN{ Hao Yin}
%    \IEEEauthorblockA{
%    \textit{haoyin@hust.edu.cn}
%    }
%    \and
%    \IEEEauthorblockN{Xiaojun Guo}
%    \IEEEauthorblockA{
%    \textit{guoxj@hust.edu.cn}
%    }
%    \and
%    \IEEEauthorblockN{ Pengyu Liu}
%    \IEEEauthorblockA{
%    \textit{eic\_lpy@hust.edu.cn}
%    }
%    \and
%    \IEEEauthorblockN{ Yayu Gao}
%    \IEEEauthorblockA{
%    \textit{yaoyugao@hust.edu.cn}
%    }
%    \and
%    \IEEEauthorblockN{ Xiaojun Hei}
%    \IEEEauthorblockA{
%    \textit{heixj@hust.edu.cn}
%    }
%    }

    \maketitle

    %\begin{abstract}
%    In this paper, a novel channel quality indicator (CQI) prediction method and an online training module are proposed for the downlink scheduling in the 5G New Radio (NR) system, aiming to reduce the impact of outdated CQI, especially in the high-speed moving situation. In the first place, we analyze the impact of outdated CQI in downlink scheduling of NR system and the recent prediction method applying in the wireless network. Then we design a data generation and online training module to test our prediction method on the ns-3 platform. The simulation result shows the proposed method has a practical effect on solving the delay of CQI on downlink transmission, and also have a better situation than feedforward neural networks
%
%\end{abstract}

    \begin{abstract}
        5G networks provide more bandwidth and more complex control to enhance user’s experiences, while also requiring a more accurate estimation of the communication channels compared with previous mobile networks. In this paper, we propose a channel quality indicator (CQI) prediction method in a deep learning approach in that a Long Short-Term Memory (LSTM) algorithm. An online training module are introduced for the downlink scheduling in the 5G New Radio (NR) system, to reduce the negative impact of outdated CQI for communication degradation, especially in high-speed mobility scenarios. First, we analyze the impact of outdated CQI in the downlink scheduling of the 5G NR system. Then, we design a data generation and online training module to evaluate our prediction method in ns-3. The simulation results show that the proposed LSTM method outperforms the Feedforward Neural Networks (FNN) method on improving the system performance of the downlink transmission. Our study may provide insights into designing new deep learning algorithms to enhance the network performance of the 5G NR system.

    \end{abstract}


    \section{Introduction}
\label{section:introduction}

In mobile communication systems, the channel condition may significantly impact the communication quality. Hence, one of the key technical advancements is to continually improve channel estimation and to enhance modulation schemes. The Channel Quality Indicator (CQI) is the most important parameter for determining the achievable data rate of multimedia transmission \cite{Rassa2018:SCOReD}. However, an ideal CQI is not always available in the base station because of the transmission delay from the user to the base station. Currently, the 5G network is still under active development, and it is designed to provide a large data rate from 100 Mbps to 1 Gbps anywhere and anytime, even in the high-speed mobile environment up to 500km/h. Unlike the LTE, the channel of the 5G network may be more dynamic and complex which makes CQI more difficult to estimate accurately. In the 5G New Radio (NR) system, the more complex and variable channel may severely throttle the throughput performance of the 5G system.

Recently, various deep learning methods have been used to solve many communication problems successfully, and many algorithms have been proposed in channel estimation and resource allocation. Ref. \cite{Yang2018:TVT} predicts channel conditions and perform channel assignments based on machine learning algorithms in wireless networks. Ref. \cite{Wu2013:JBUPT} and \cite{Abdulhasan2014:ISTT} proposed to apply the Feedforward Neural Network (FNN) in predicting the CQI based on the historical CQIs in LTE. However, the evaluation experiments were built based on simplified proprietary simulations. In this paper, we are motivated to study how to improve the accuracy of the CQI prediction for the downlink scheduling on base stations to fulfill the requirements of the 5G network based on a transparent open-source simulation framework.

Ns-3 is a popular open-source network simulation tool for the system-level simulation. Over the years, ns-3 has been instrumenting the 5G protocol stack with sufficient technical details for high fidelity at different network layers. For the 5G NR network simulation, there has been the newly released millimeter-wave module \cite{Mezzavilla2018:COMST} and the NR module \cite{Bojovic2018:wns3}. In this paper, we propose to use ns-3 to generate simulation data for training the deep learning algorithms and conduct a performance comparison study between different CQI prediction algorithms with the same system-level simulation settings. Besides, an interface between these two frameworks is required to enable the data exchange from the network simulator to popular AI frameworks. We apply the Long Short-Term Memory (LSTM) model to design the CQI prediction method and also apply FNN as a comparison benchmark. The specific contributions of this work include:
\begin{itemize}
    \item Design the data generation and the online training module in ns-3 which enables repeatable evaluation experiments for various deep learning methods in open-source network simulations.
    \item Propose an LSTM-based CQI prediction method to improve the CQI prediction accuracy than Feedforward Neural Networks.
\end{itemize}

This paper is organized as follows. In Section \ref{section:problem}, we present the downlink scheduling problem in the 5G NR system. Then, we propose an LSTM method to predict CQI in Section \ref{section:method}. In Section \ref{section:results} we report the ns-3 simulation results for performance comparison. Finally, we conclude the paper in Section \ref{section:conclusion}.

    \section{Problem Statement}
\label{section:problem}
\begin{comment}
\begin{table}
    \centering
    \caption{Notation}
    \label{table:define}
\begin{tabular}{|c|c|}
    \hline
    $R_t$&throughput\\
    \hline
    $n$&the UE number\\
    \hline
    $M_{i,t}$&the MCS for $i^{th}$ UE at time $t$\\
    \hline
    $r(M_{i,t})$&the obtained transmission rate\\
    \hline
    $B_{i,t}$&the block error rate of the transport block\\
    \hline
    $E_p$&prediction error\\
    \hline
    $E_t$&time delay error\\
    \hline
    $E(i)$&time delay error or prediction error at time i\\
    \hline
    $CQI_{pred}(t)$&predict CQI at time $t$\\
    \hline
    $CQI_{real}(t)$&real CQI at time $t$\\
    \hline
    $N$&the length of data used for each prediction\\
    \hline
    $K$&the length of input data\\
    \hline
    $\tau$&all the delay of CQI arriving\\
    \hline
    $MSE$&mean-square error\\
    \hline
    \end{tabular}
\end{table}

Table \ref{table:define} shows all the notation may be used in this paper.
\end{comment}

Evolving from the LTE system, the 3GPP/NR system supports multiple different types of subcarrier spacing as NR numerology in the time domain. The basic frame structure of NR is granular with slots, and when divided downwards, the number of slots included in each subframe can be flexibly set according to the subcarrier spacing, thereby providing a flexible and reliable extension. In this case, a frame can be divided into several slots, which enables scheduling decisions made in a shorter time known as Transmission Time Interval (TTI) \cite{3gpp.38.211}. In the frequency domain, each Resource Block (RB) consists of 12 subcarriers normally and is expected to experience specific propagation and interference conditions. These conditions are evaluated by the user side and reported to the base station (BS) as the channel quality indicator (CQI), which allows the BS to take advantage of the frequency diversity of the radio channel.

Generally, in OFDM access used by NR and LTE systems, the implementation of high spectral efficiency requires dividing the total bandwidth into a plurality of narrow sub-bands for each user to share. To maximize the spectral efficiency, Modulation and Coding Scheme (MCS) is adopted, which requires the system to know the channel quality in time and accurately and then make a reasonable MCS selection. Whether the scheduler can obtain channel quality indicator (CQI) with accurate real-time feedback is the key to achieving high scheduling efficiency. 

In the NR system, a downlink scheduler (MAC Scheduler) is located at a medium access (MAC) layer and receives the scheduling information provided by each user accessed to the base station, like CQI, buffer status report (BSR) sent by the Radio Link Control (RLC) layer, Quality of Service (QoS), etc. Then the scheduler selects a user for each RB according to the information mentioned above. As the RBs are allocated to the users, the MCS will be calculated by the CQI value, further determining the Transport Block Size (TBS). The MCS and TBS determine how much data will be transmitted at this time slot. The CQI is reported by the users at each scheduling time. Both the request from the base station and the transmission of CQI will cause a delay when the BS receives the CQI feedback \cite{Rassa2018:SCOReD}. Under ideal conditions (CQI feedback delay is 0), the throughput and MCS at time $t$ in the BS is shown is Equ. (\ref{eq:thro_ori}). Where $n$ is the UE number, $M_{i,t}$ is the MCS for $i^{th}$ UE at time $t$, $r(M_{i,t})$ is the obtained transmission rate determined by the given MCS $M_{i,t}$, and $B_{i,t}$ is the block error rate of the transport block.
\begin{equation}
   R_t = \sum_{i=1}^{n}[r(M_{i,t})(1-B_{i,t})] \label{eq:thro_ori}
\end{equation}

However, in the real network, the CQI feedback has a delay which leads to the scheduling use the $CQI(t-\tau)$ instead of the $CQI(t)$. Therefore, the actual MCS changes to $M_{i,t-\tau}$ and the actual data rate $R_t^{true}$ at time $t$ is shown in Equ. (\ref{eq:thro_dealy}).
\begin{equation}
R_t^{true} = \sum_{i=1}^n [r(M_{i,t-\tau})(1-B_{i,t}] \label{eq:thro_dealy}
\end{equation}

\begin{figure}[htb]
   \centering
   \includegraphics[width=.5\textwidth]{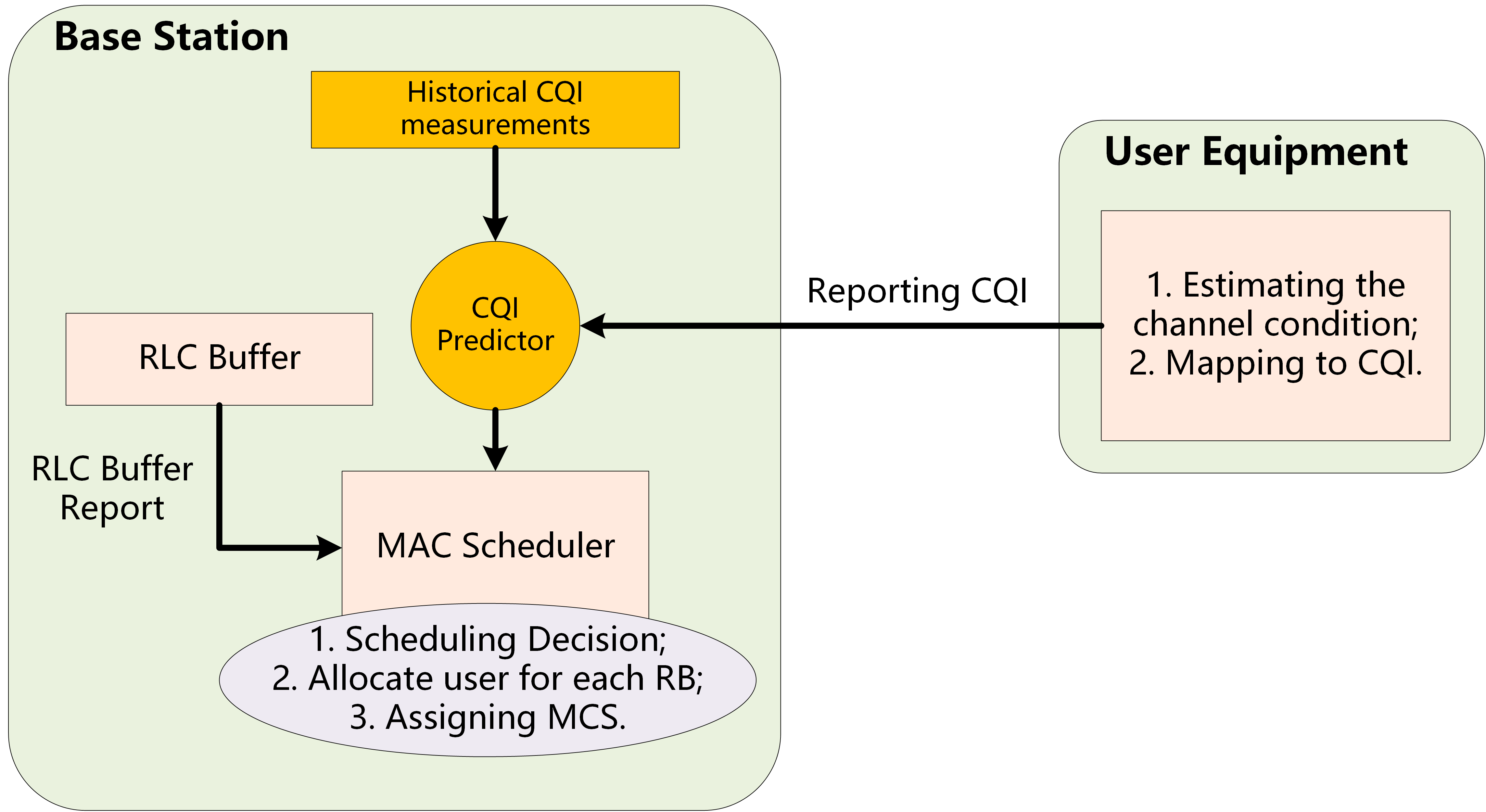}
   \caption{Illustration of CQI prediction module.}
   \label{fig:cqi}
\end{figure}

In this specific time, $t$, on the one hand, if the $CQI(t-\tau)$ is larger than the real value, but the channel condition is not as well, the error block rate will increase. On the other hand, if the $CQI(t-\tau)$ is smaller, which leads to the limitation of MCS, the transmission of packets will decrease. In both situations, it will cause a decrease in throughput. Especially in the high-speed moving situation like in high-speed rail, the changing and varying channel condition makes this problem more severe. Comparing with the LTE system, one of the key technologies of the NR system is the millimeter-wave channel used for data transmission. One feature of the millimeter-wave channel is that it has a large attenuation in air and a weak diffractive power, so it is more important but harder to get more reliable CQI values. Thus, in this paper, we propose a CQI prediction method to improve the CQI accuracy in downlink scheduling in the NR system. The integrated CQI prediction method with the downlink scheduling module is shown in Fig. \ref{fig:cqi}.

    \section{Method}
\label{section:method}

In the 5G network, the physical layer resources are more abundant but more complicated than LTE. Consequently, more flexible and reliable scheduling algorithms and more accurate CQI values are of great significance for improving NR networks. Traditionally, the CQI generation and reporting are like to cause a delay in the scheduling that will result in a drop in system performance. Using prediction methods like deep learning algorithms can improve the accuracy of the CQI, which finally benefits the NR system.

\subsection{Using Deep Learning Algorithms to Predict CQI}
Currently, we focus on the wideband CQI, a positive integer from 1 to 15. As the CQI is only a discrete value and UEs are changing from time to time, it's not possible to simply train a module to predict all users' behavior with just the history value of CQI. If a user is moving toward the base station at this time, while another time it is moving away from the BS, the module will be confused about how the user will behave, and then it learns merely to imitate the past behavior. In this way, the training module will fit after only a few steps, and the result shows an apparent delay of the CQI result comparing to the real value reported. What's more, the behavior of users is varied, so the BS should maintain different models for different users attached to it.

Although a well-trained module is troublesome to utilize in the prediction of CQI, a user's action and the channel condition would not change in a short time, which means it is practical to predict the CQI value in a time period. When the state changes and the prediction accuracy decreases rapidly, a new module will be trained again to adapt to the new features.

\subsubsection{Online Training Method}
When the MAC layer starts to schedule the users for downlink transmission, it handles the CQI reported from the users. The time consuming of the generation and transmission of CQI causes the base station to use a delayed CQI value $CQI_{delay}(t)$ for scheduling (using $\tau$ to represent the delay of CQI arriving, then $CQI_{real}(t-\tau)$ = $CQI_{delay}(t)$) We'd like to use the $CQI_{pred}$ predicted by the deep learning algorithm to replace the $CQI_{delay}(t)$.
 
 An online learning model is introduced to perform a timely update on the prediction because the distribution of CQI sequences changes with time in the wireless channel. At the beginning of the downlink transmission for a specific user, a pre-trained model will be trained and adjusted by using the CQI values reported from a specific user. When the users are attached to the BS, the BS will maintain the CQI prediction model for each user. After several steps, the neural network could learn the current features of the channel state then provide a reliable prediction. Then the BS will stop training to reduce the consumption of the resources. After a while, the channel condition may change to another way due to the environment or behavior of the users, causing the prediction accuracy dropped rapidly, and the model will restart to train with the latest date to converge again. The decision of training or not is based on the comparison of weighted mean square error (MSE) in a time window $K$. For each scheduling, we calculate the the error for both using $CQI_{delay}(t)$ and $CQI_{pred}$. After the real CQI at the time $t$ reported from users arrived, the prediction error $E_p$ and time delay error $E_t$ are calculated by following Eq. (\ref{eq:ep}) and Eq. (\ref{eq:et}).
\begin{equation}
    E_p(t) = CQI_{pred}(t) - CQI_{real}(t)\label{eq:ep}
\end{equation}

\begin{equation}
    E_t(t) = CQI_{real}(t-\tau) - CQI_{real}(t)\label{eq:et}
\end{equation}

Based on a series of CQI values in the windows size $K$ from current time, BS calculates the weighted MSE using Eq. (\ref{eq:sum_error}). The different error values in this window have different weights on this evaluation, the closest one to the current time has lager weight while the further one has less weight. After the latest real CQI value is received, the system updates the weighted MSE both for the prediction result and time delay result. If the $MSE_{pred}$ is smaller or equal to the $MSE_{delay}$, it means in the current time series, the prediction result is better, so we still stop training and use the current model to predict the CQI. Otherwise, when the $MSE_{pred}$ is greater, the user behavior is tended to change, so the $CQI_{real}(t-\tau)$ will be used for scheduling and a new module will be trained to match the new feature until the prediction error is smaller than the delay value.  
\begin{equation}
\begin{aligned}
    MSE_{delay}{(t)} = \sum_{i=t-K}^{t}\left(E_t(i)^{2} \cdot \frac{i+1-(t-K)}{K}\right)\\
    MSE_{pred}{(t)} = \sum_{i=t-K}^{t}\left(E_p(i)^{2} \cdot \frac{i+1-(t-K)}{K}\right)\label{eq:sum_error}
\end{aligned}
\end{equation}

\subsubsection{Using LSTM Predict CQI}

\begin{figure}[htb]
    \centering
    \includegraphics[width=.5\textwidth]{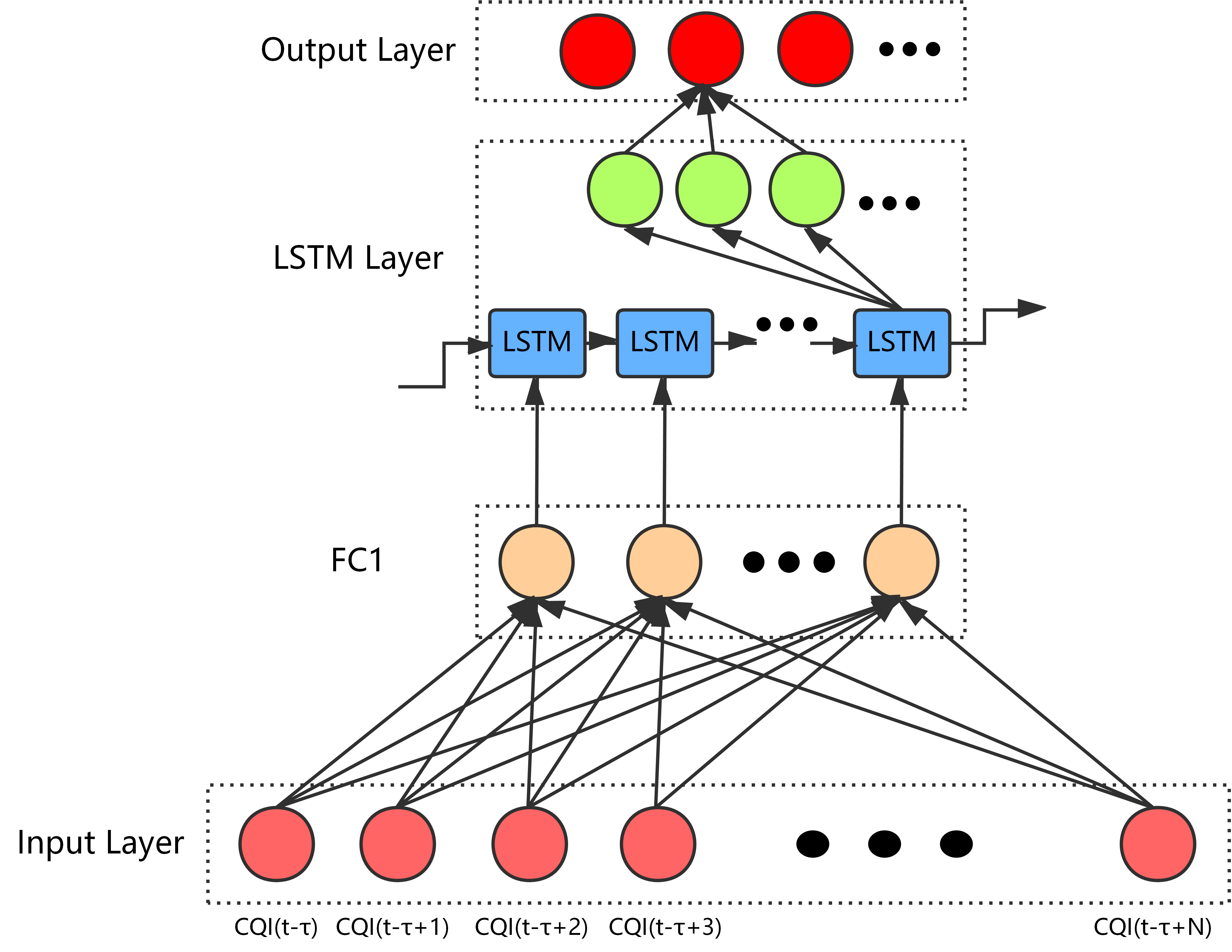}
    \caption{LSTM architecture.}
    \label{fig:lstm}
\end{figure}

Long Short-Term Memory (LSTM) is ideal for dealing with issues that are highly correlated with time series, and the changing of CQI in the BS side can be evaluated and predicted by the past series of CQI. So we apply LSTM as the core network to predict the CQI value according to the historical information. The key function of LSTM is that it has a certain memory effect which can remember the history feature of the CQI and then using this information to predict the future CQI — shown in Fig. \ref{fig:lstm}, the module is consists of the fully connected layer, the LSTM layer, the prediction layer, and an update module. The fully connected layer is connected for feature extraction, and then the LSTM layer is implemented to extract temporal information. Finally, the prediction layer will sum the vector from the output of the LSTM layer, which produces the predicted value of CQI. The input layer is a vector of CQI from $CQI_{real}(t-\tau -N)$ to $CQI_{real}(t-\tau)$, $N$ stands for the length of data used for each prediction. $N_{FC}$ neural units are used for the fully connected layer and the output of the module is a vector of length $N_{LSTM}$.

$Selu$ is selected as the activation function, and because it can guarantee a smooth gradient in the training\cite{Klambauer2017:NIPS}. Considering CQI is a series of discrete numbers, the training data has a larger interval, which may lead to a large gradient causing the death of the neural units in other ways. Besides, as implementing the online training method, the larger learning rate may cause the parameter change from time to time easily that necessitates a guarantee of the output of gradient.

\subsubsection{Other Deep Learning Method}

Feedforward neural networks (FNN) mentioned above is also implemented as the prediction module. The online training method is applied to it. The FNN is consists of an input layer, a hidden layer, and a prediction layer. Unlike LSTM, the FNN won't consider the time as a feature of data. The input of the FNN is a CQI vector of length $nN$, and the dimension of the hidden layer is $N_{H}$. Other settings are the same as LSTM.

\subsection{Data Generation and Interaction Between ns-3 and AI Frameworks}
Although many recent works started to utilize deep learning algorithms to the research of wireless networks, it is based on some link-level simulation rather than a whole system-level simulation. It may cause some limitations and inaccuracies due to the link-level simulation only considering a little part of the network. Besides, it is also hard to generate data and enable real-time tests of the algorithm. To conquer these problems, we develop a data interaction module between ns-3 and AI frameworks to generate data and allow online training and testing methods.

\subsubsection{Data Generation}

We promote a new way to generate data from open-source network simulation tools like ns-3. These open-source tools are often used in research and education and are maintained by many experienced developers. Besides, we can analyze the simulator directly from the source code and can set different scenarios to verify the simulator and test the data reliability. It is also easy and convenient to generate all the kinds of data we need and save for the learning algorithms. Finally, we can use the same type of data for further experimental verification in the simulator.

In this paper, the ns-3 NR module \cite{Patriciello2019:wns3} is adopted to generate CQI data and build system-level simulation to compare the performance of the different approaches. The CQI is generated on the UE side and received on the BS side in ns-3. In the UE side, after the reception of data packets, the PHY layer calculates the wideband SINR taking the path loss into account, MIMO beamforming gains, and frequency-selective fading. This triggers the generation of CQI reports, which are fed back to the base station in either UL data or control slots.In ns-3 framework, there are two modules to calculate the DL-CQI,  \textsl{PiroEW2010} and \textsl{MiErrorModel}. These two ways of calculating CQI are both based on the 4-bit CQI modulation table by calculating the SINR and spectral efficiency to get the final value of CQI. The factors affecting SINR are the distance of UEs, the speed of UEs, and the types of scenarios. So, in general, CQI is determined by the conditions of UEs, the distance and the speed, and the different scenarios. Since \textsl{MiErrorModel} is widely used in LTE simulation, we pick \textsl{MiErrorModel} to generate data.

The process to calculate CQI using \textsl{MiErrorModel} is based on Bit Error Rate(BER). At the BS side, the BS can calculate the BER of UEs through their SINR, and evaluate the MCS of the UEs according to the BER to obtain the optimal transmission performance of the UEs when BER is less than 10\%. For the calculation of SINR, the traditional way is shown as the Eq. (\ref{eq:tra_SINR}).
\begin{equation}
    SINR = \frac{Power_{Signal}}{Power_{Interfere_Noise}}\label{eq:tra_SINR}
\end{equation}

 As the bandwidth for signal and interference is the same so the SINR can be calculated as the modified Eq. (\ref{eq:ns_SINR}), whose PSD stands for power spectral density.

\begin{equation}
    SINR = \frac{RxPsd}{Psd_{InfNoise}}\label{eq:ns_SINR}
\end{equation}

 ${RxPsd}$ means the power spectral density of the RX signal, while  ${Psd_{InfNoise}}$ is the power spectral density of interference. The  ${RxPsd}$ can be calculated by ${TxPsd}$, which is the power spectral density for transmission in BS.

\subsubsection{Interaction Module}
\begin{figure}[htb]
    \centering
    \includegraphics[width=.5\textwidth]{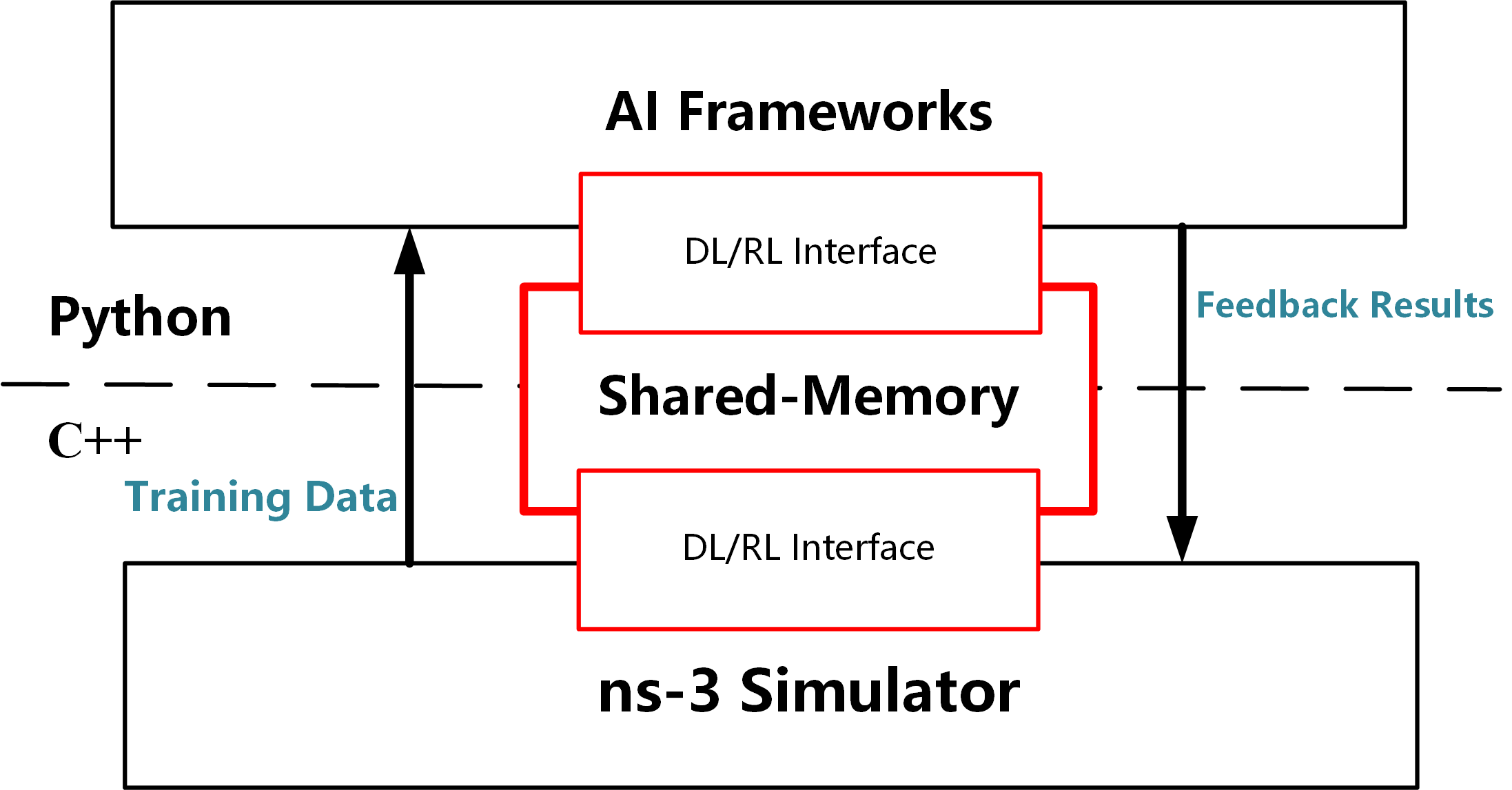}
    \caption{Interaction module.}
    \label{fig:ns3-ai}
  \end{figure}

The ns-3 simulator and AI frameworks are running in different binary and hard to integrate with each other. The key function is to implement a new module to exchange the important data required by both frameworks. The ns-3 provides the data as the input of the deep learning algorithm to train the model and holds itself until getting the feedback data from the DL algorithms. In this way, it only transfers the data from both sides without having any impact on the original simulation process, which looks like simply adding an algorithm directly into the simulator. The interaction module should be fast and reliable to fulfill the request of simulation. Thus we used shard-memory to enable the data interaction to reduce the data transferring time and increasing the reliability with readers–writer lock in python. The interaction module between ns-3 and AI frameworks is shown in Fig. \ref{fig:ns3-ai}.

 In this module, users can exchange data between ns-3 and AI frameworks by putting the data in the same shared-memory.  A locking method is introduced in the integration module to avoid the conflict of operating the same memory by updating its version after writing in new data to notify the consumers. The data generated by ns-3 directly maintains in the shared-memory, so there is no memory copy that reduced the consumption of both memory and time in the simulation.  

    \section{Experiment Results}
\label{section:results}
In this section, we evaluate the performance of the CQI prediction based on two basic scenarios utilizing our online training method in the ns-3 NR module. Both the prediction accuracy and the impact on throughput are measured and analyzed.

\subsection{Simulation Scenarios}
\begin{figure}[] 
  \centering
  \subfigure[Scenario 1: Single user. ] {\includegraphics[width=.2\textwidth]{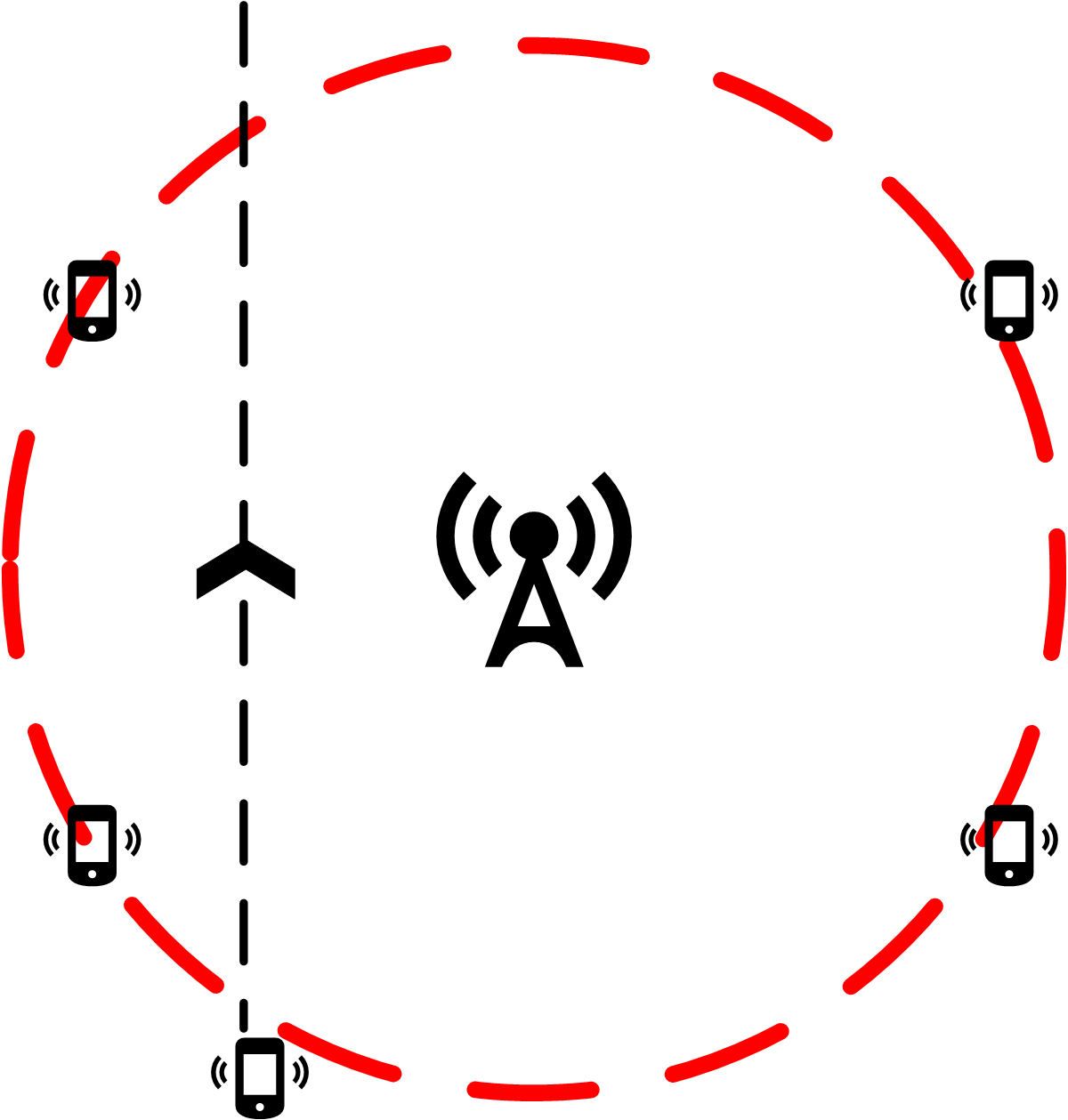}}\quad
  \subfigure[Scenario 2: Multi-user.] {\includegraphics[width=.22\textwidth]{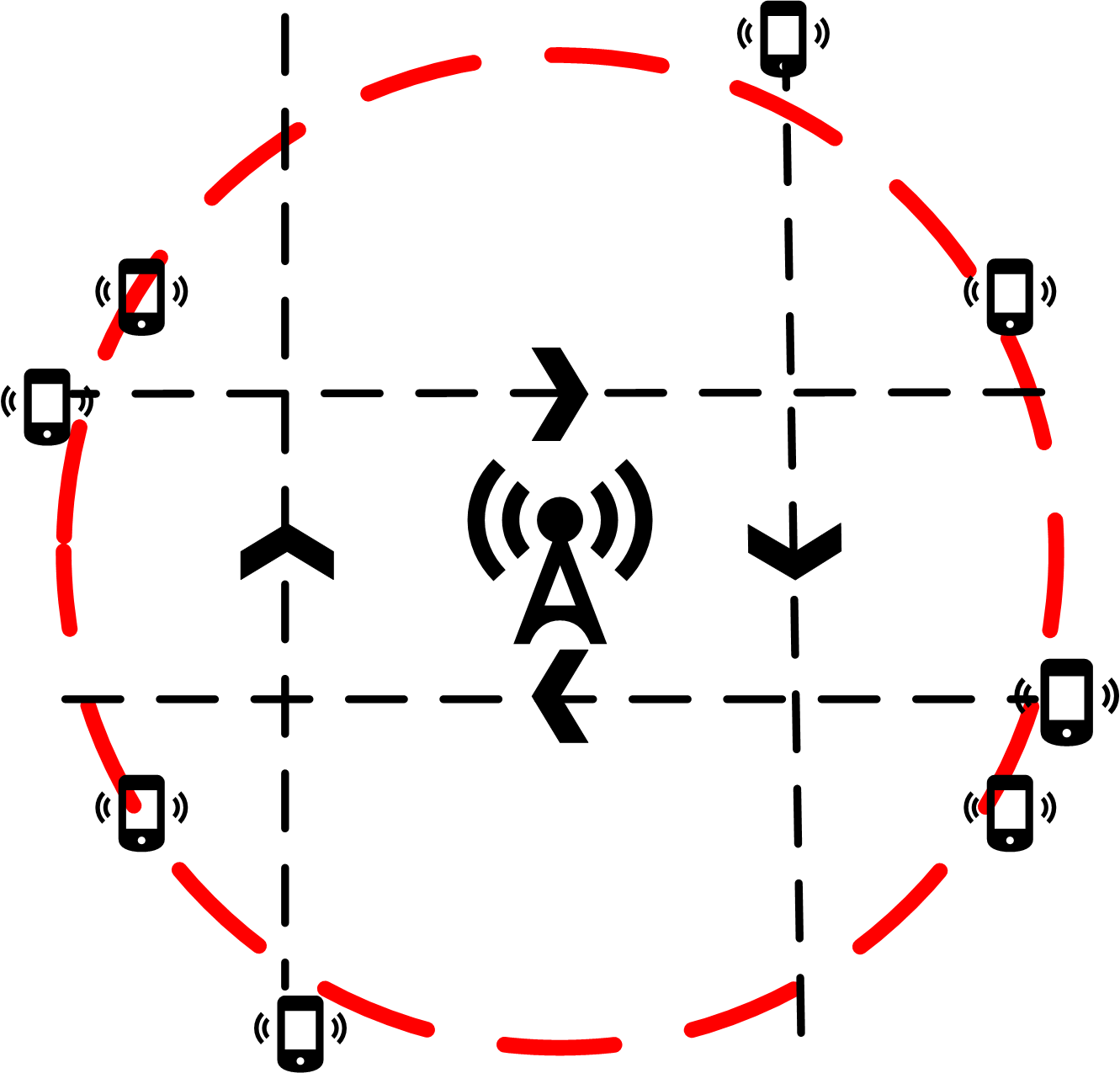}}
  \caption{Graphic illustration of simulation scenarios.}
  \label{fig:scene}
\end{figure}

The ns-3 simulation platform is adopted for the system-level simulation and data generation. We consider two scenarios to estimate the performance of the prediction method. In the first simulation scenario, four users are attached to the base station as the interfering users. Another user marked as the moving user is crossing the base station from outside the coverage to another side. The speed of the moving user varies in different scripts, which provides us with different results. In the second scenario, based on the first one, four users moving fast in a different direction to evaluate the performance in the multi-user scenario. Both scenarios are using the same parameter and settings. The millimeter-wave channel module is utilized, and numerology is setting to 1 to reduce the simulation time. The delay $\tau$ of the CQI using by BS  is set to 0.5 ms to provoke a delay simply equal to one slot. The simulation scenario is shown in Fig. \ref{fig:scene} and the parameters are shown in table \ref{table:sim_pra}. The CQI value of the interfering users won't change a lot because they keep still during the simulation. We apply the deep learning method for each moving user, and the parameters used for the training module are shown in table \ref{table:lstm}.

\begin{table}
    \centering
    \caption{Simulation Parameters}
    \label{table:sim_pra}
    \begin{tabular}{|c|c|}
    \hline
    \textbf{Parameter} & \textbf{Value}\\ \hline
    Power& 20 dBm \\\hline
    Numerology& 1 \\\hline
    Bandwidth&  100MHz  \\\hline
    Scenario& UMa \\  \hline
    Data Rate& 64Mbps\\  \hline
    RLC Mode &UM\\  \hline
    Moving Module &Constant Speed \\ \hline
    Speed   &10m/s 20m/s 30m/s \dots\\  \hline
    Scheduler Algorithm   &Random Robin\\  \hline
    \end{tabular}
\end{table}

\begin{table}
    \centering
    \caption{Training Module Parameters}
    \label{table:lstm}
    \begin{tabular}{|c|c|}
    \hline
    \textbf{Parameter} & \textbf{Value}\\ \hline
    Batchsize& 20 \\\hline
    Pre\_train Times& 20 \\\hline
    LSTM Unit&  30  \\\hline
    Input Length (K)& 200 \\  \hline
    Data Length Using for Prediction (N)& 40\\  \hline
    Activation Function & Selu\\  \hline
    \end{tabular}
\end{table}

If the users are not moving fast or changing their positions rapidly, the CQI will keep the same for a long time, so the prediction results are quite the same as using the outdated CQI. In this paper, the fast-moving scenario is our first concern about the CQI prediction. From the discussion above, we know that CQI will affect the throughput of each user, therefore besides the CQI prediction accuracy, we also integrate our online training module with ns-3 to evaluate the performance with the real-time changing of CQI.

For system performance, we use Random Robin as the MAC scheduler, which allocates RB to each user fairly, so that the scheduling algorithm would not impact the throughput of the specific user. Therefore, we consider the specific moving users' throughput as the performance of the system because others' throughput tends to be the same. 

\subsection{Results}
\subsubsection{CQI Prediction}

\begin{figure}[htbp] 
  \centering
  \subfigure[Scenario 1: Single user prediction accuracy. ] {\includegraphics[width=.4\textwidth]{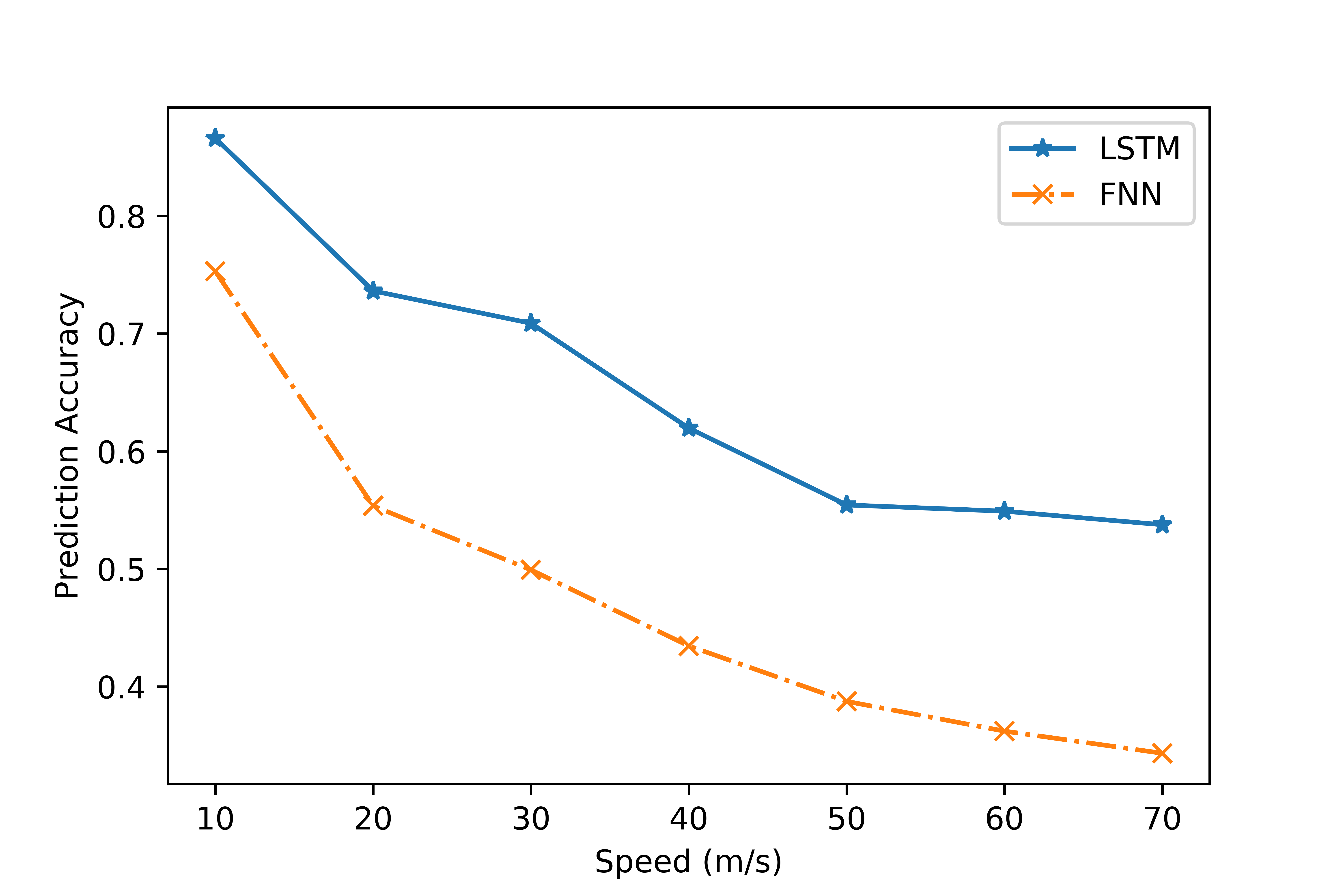}}\quad
  \subfigure[Scenario 2: Multi-user average prediction accuracy.] {\includegraphics[width=.4\textwidth]{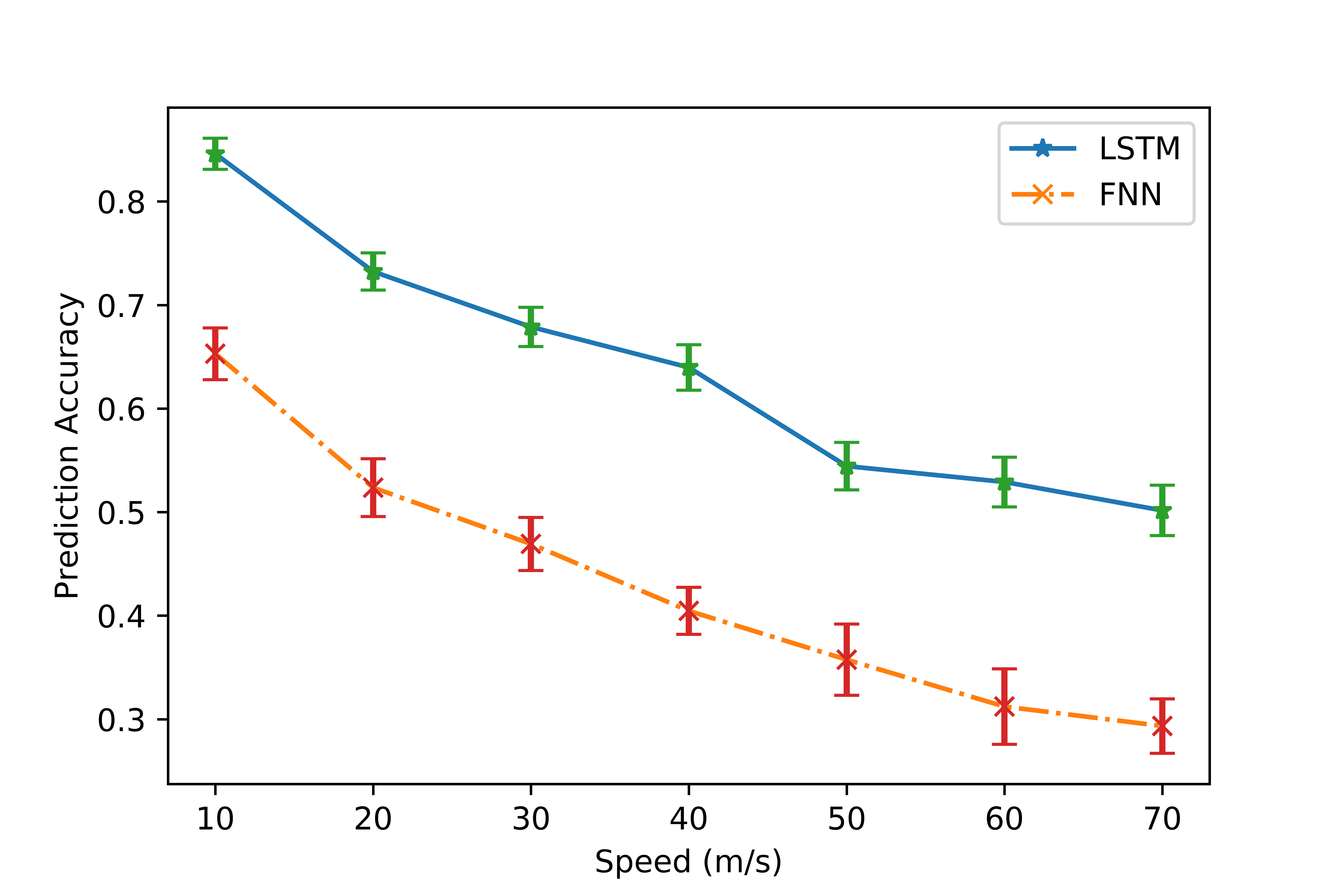}}
  \caption{Prediction accuracy versus speed.}
  \label{fig:usage}
\end{figure}

Using the data generated from ns-3, we test the accuracy of different scenarios, and the result is shown in Fig. \ref{fig:usage}. Two different prediction methods are applied as the prediction module, the FNN and LSTM. Form the figures, as the speed improved, the CQI is more difficult to predict, so the total prediction accuracy is down. The LSTM module is much more reliable than the FNN module, which can have accuracy above 50\% while the speed of the users increased to 70 m/s. As the speed is increasing, the prediction accuracy is decreasing, and the FNN could not provide a reliable prediction while LSTM is still working. From the multi-user scenario, we use the average prediction accuracy to evaluate the performance. The error bars represent one standard deviation of the mean value to estimate the difference between different users. In general, the errors of the average are increasing as speed increases, but the errors of LSTM is smaller than FNN. Thus the performance of LSTM is also better in the multi-user scenario.

\subsubsection{System Performance}
\begin{figure}[htbp] 
  \centering
  \subfigure[Scenario 1: Single user throughput. ] {\includegraphics[width=.4\textwidth]{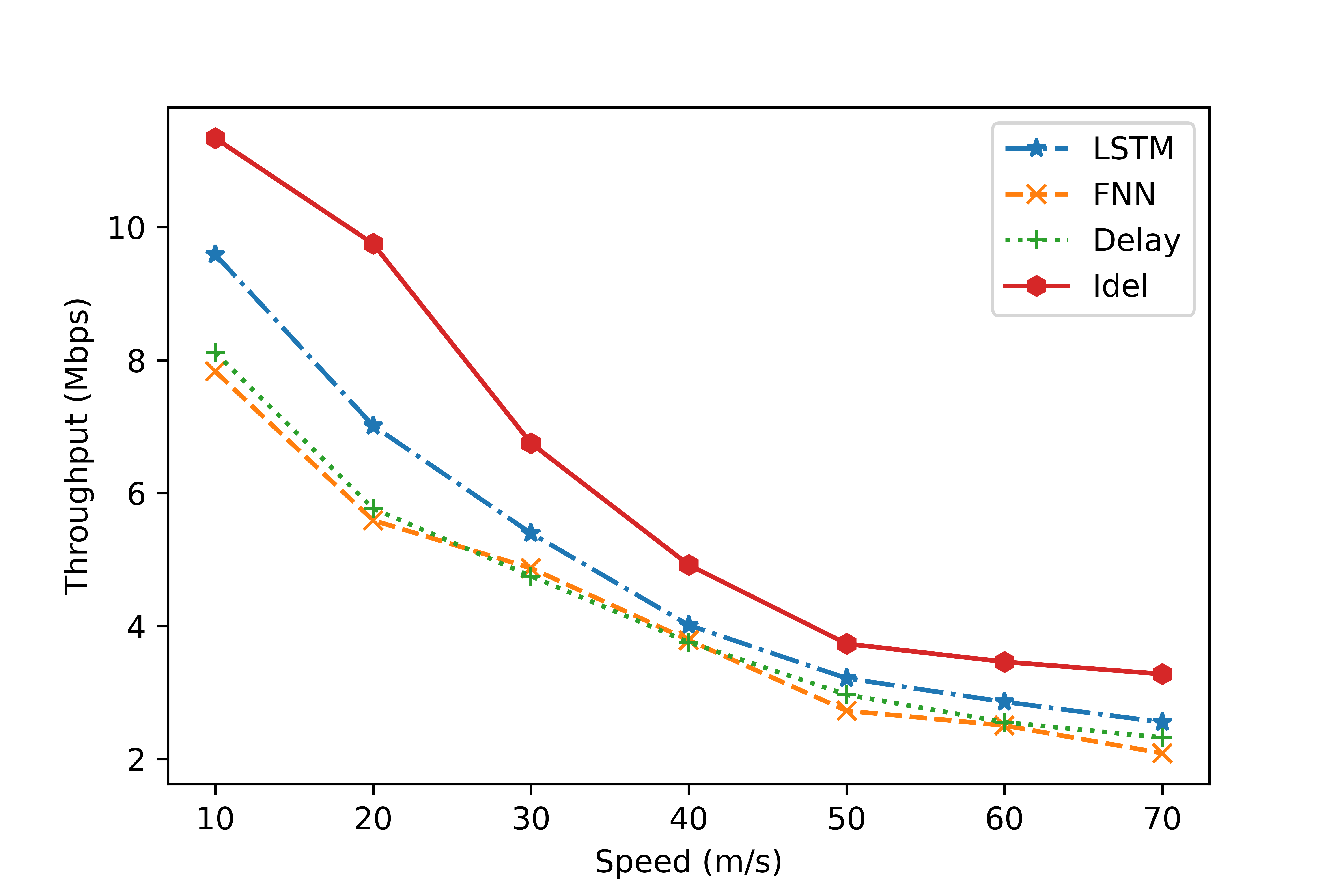}}\quad
  \subfigure[Scenario 2: Multi-user average throughput.] {\includegraphics[width=.4\textwidth]{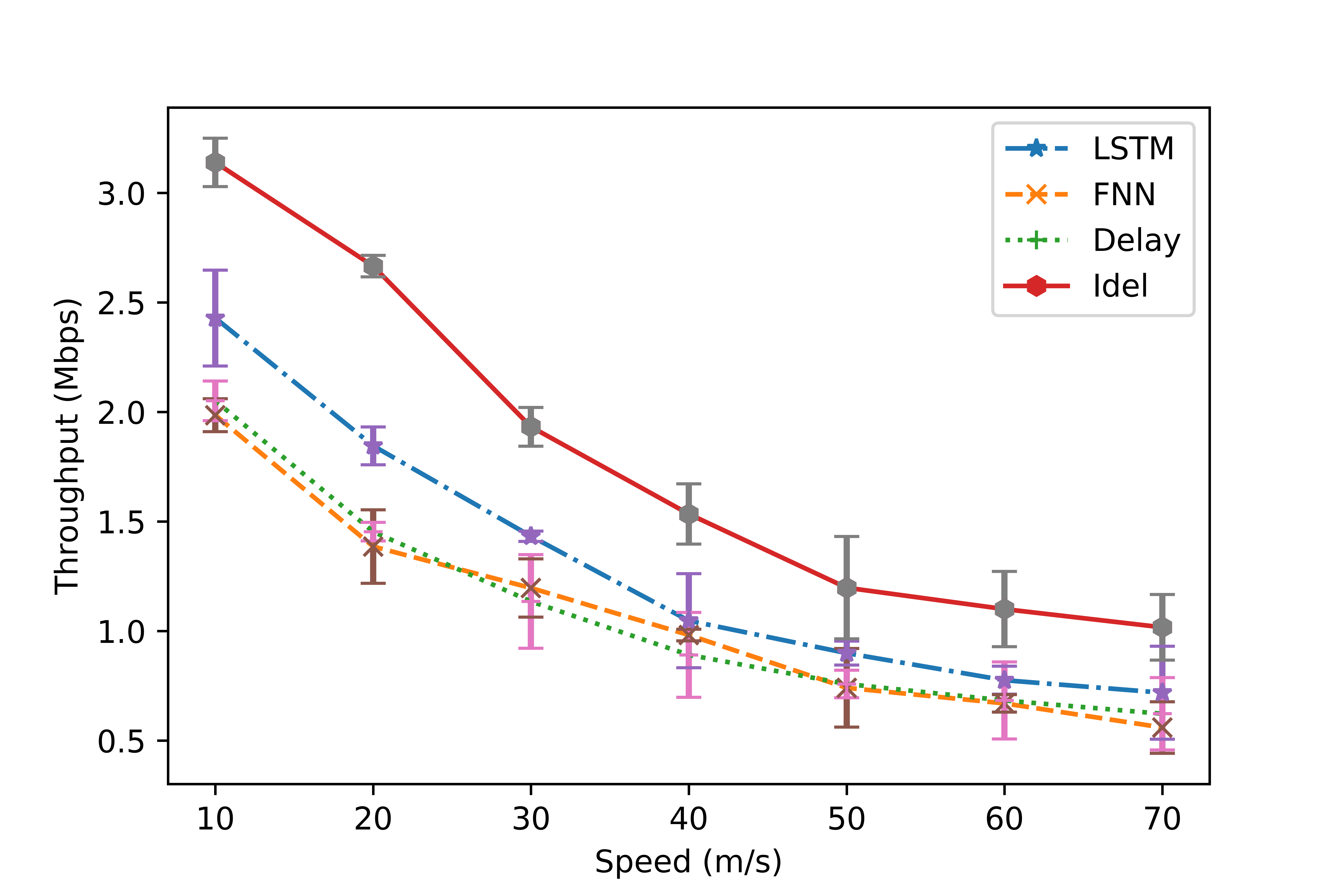}}
  \caption{Throughput versus speed.}
  \label{fig:throughput}
\end{figure}

For the system level performance test, we integrate the online training module in the ns-3, so that our predicted CQI could be used directly in ns-3. The result of the throughput changing versus different speeds is shown in Fig. \ref{fig:throughput}. We compare the four methods of the CQI reporting method on the impact of throughput in the simulation. From the simulation results, both in the single-user and multi-user scenarios,   the LSTM performed better than FNN and delayed CQI. In the multi-user scenario, we obtain a deep learning model for each user, so it won't predict by a signal model that may be confused by the moving direction of different users. In this way, the prediction method still works in the multi-user scenario that can provide an improvement in the performance.

\balance

\section{Conclusion}
\label{section:conclusion}

For the adversarial effects of outdated CQI on the 5G NR downlink scheduling, in this paper we propose a LSTM-based CQI prediction method and design an online training module to improve the accuracy of the CQI prediction. The simulation results show that in a high-speed mobility scenarios, our proposed method outperforms FNN in terms of prediction accuracy and the throughput performance. In the future, we plan to conduct more simulation experiments with high bandwidth and numerology settings to evaluate the performance in high transmission speeds and millimeter-wave channels. We will also design other deep learning methods to improve the CQI prediction accuracy with less complexity. In addition, we plan to implement the proposed algorithms in our instrumented 5G NR testbed to examine the technical feasibility with the computation cost of the deep learning based prediction methods for real base stations.

    \bibliographystyle{IEEEtran}
    \bibliography{reference}

\begin{thebibliography}{1}
\providecommand{\url}[1]{#1}
\csname url@samestyle\endcsname
\providecommand{\newblock}{\relax}
\providecommand{\bibinfo}[2]{#2}
\providecommand{\BIBentrySTDinterwordspacing}{\spaceskip=0pt\relax}
\providecommand{\BIBentryALTinterwordstretchfactor}{4}
\providecommand{\BIBentryALTinterwordspacing}{\spaceskip=\fontdimen2\font plus
\BIBentryALTinterwordstretchfactor\fontdimen3\font minus
  \fontdimen4\font\relax}
\providecommand{\BIBforeignlanguage}[2]{{%
\expandafter\ifx\csname l@#1\endcsname\relax
\typeout{** WARNING: IEEEtran.bst: No hyphenation pattern has been}%
\typeout{** loaded for the language `#1'. Using the pattern for}%
\typeout{** the default language instead.}%
\else
\language=\csname l@#1\endcsname
\fi
#2}}
\providecommand{\BIBdecl}{\relax}
\BIBdecl

\bibitem{Rassa2018:SCOReD}
E.~H.~R. {Rassa}, H.~A.~M. {Ramli}, and A.~W. {Azman}, ``Analysis on the impact
  of outdated channel quality information ({CQI}) correction techniques on
  real-time quality of service ({QoS}),'' in \emph{IEEE Student Conference on
  Research and Development ({SCOReD})}, Nov 2018.

\bibitem{Yang2018:TVT}
S.~{Yang}, Y.~{Tseng}, C.~{Huang}, and W.~{Lin}, ``Multi-access edge computing
  enhanced video streaming: Proof-of-concept implementation and
  prediction/{QoE} models,'' \emph{IEEE Transactions on Vehicular Technology},
  vol.~68, no.~2, pp. 1888--1902, Feb 2019.

\bibitem{Wu2013:JBUPT}
B.~{Wu}, Z.~{Mi}, W.~{Wang}, Y.~{Lu}, and Z.~{Zhu}, ``A {LTE} downlink
  scheduling based on {CQI} predicted by neural network,'' \emph{Journal of
  Beijing University of Posts and Telecom}, 2013.

\bibitem{Abdulhasan2014:ISTT}
M.~Q. {Abdulhasan}, M.~I. {Salman}, C.~K. {Ng}, N.~K. {Noordin}, S.~J.
  {Hashim}, and F.~B. {Hashim}, ``A channel quality indicator ({CQI})
  prediction scheme using feed forward neural network ({FF-NN}) technique for
  {MU-MIMO} {LTE} system,'' in \emph{IEEE International Symposium on
  Telecommunication Technologies ({ISTT})}, Nov 2014.

\bibitem{Mezzavilla2018:COMST}
M.~{Mezzavilla}, M.~{Zhang}, M.~{Polese}, R.~{Ford}, S.~{Dutta}, S.~{Rangan},
  and M.~{Zorzi}, ``End-to-end simulation of {5G} {mmWave} networks,''
  \emph{IEEE Communications Surveys \& Tutorials}, vol.~20, no.~3, pp.
  2237--2263, 2018.

\bibitem{Bojovic2018:wns3}
B.~Bojovic, S.~Lagen, and L.~Giupponi, ``Implementation and evaluation of
  frequency division multiplexing of numerologies for {5G} new radio in ns-3,''
  in \emph{Proceedings of the 10th Workshop on Ns-3}, 2018.

\bibitem{3gpp.38.211}
3GPP, ``{NR; Physical channels and modulation},'' Tech. Rep., Jan 2018.

\bibitem{Klambauer2017:NIPS}
G.~Klambauer, T.~Unterthiner, A.~Mayr, and S.~Hochreiter, ``Self-normalizing
  neural networks,'' in \emph{Advances in Neural Information Processing
  Systems}, 2017.

\bibitem{Patriciello2019:wns3}
N.~Patriciello, S.~Lagen, L.~Giupponi, and B.~Bojovic, ``An improved {MAC}
  layer for the {5G} {NR} ns-3 module,'' in \emph{Proceedings of the 2019
  Workshop on Ns-3}.

\end{thebibliography}
%    \vspace{12pt}

    \end{document}